\documentclass{iau_FM}

\usepackage[T1]{fontenc}
\usepackage{amsmath}
\usepackage{graphicx}
\usepackage{multirow}
\usepackage{enumitem}
\setlist[description]{leftmargin=\parindent,labelindent=\parindent}

\title[Indigenous rights and outer space ] 
{Overview of Indigenous rights and outer space for the IAU-CPS Policy Hub}

\author[Hilding Neilson]   
{Hilding Neilson$^1$
}

\affiliation{$^1$Department of Physics \& Physical Oceanography, \\ Memorial University of Newfoundland \& Labrador,
St.~John's, NL A1B 3X7, Canada\\ email: {\tt hneilson@mun.ca} \\[\affilskip]
}

\pubyear{2024}
\jname{Astronomy in Focus, Volume 1} 
\editors{Piero Benvenuti, ed.}
\begin{document}

\maketitle

\begin{abstract}
As part of the mission of the International Astronomical Union Centre for the Protection of the Dark and Quiet Sky from Satellite Constellation Interference (IAU-CPS) Policy Hub to consider national and international regulations about the usage and sustainability in outer space, we also included discussion specific to the rights of Indigenous peoples with respect to outer space under the context of the United Nations Declaration for the Rights of Indigenous Peoples (UNDRIP).  In this work, we review how some of the articles of UNDRIP require various actors in the use and exploitation of outer space including satellite companies, nation states, and professional/academic astronomy to consult and support Indigenous peoples/nations and respect Indigenous sovereignties.  This work is concluded with recommendations for consulting and collaborating with Indigenous peoples and recommendations for moving from the traditional colonial exploitation of outer space and building an anti-colonial future in relationship with outer space.

\end{abstract}

\firstsection 
\section{Introduction}

The purpose of this section is to explicitly highlight Indigenous rights and perspectives with respect to outer space and to consider the importance of Indigenous methods, consultations, and sovereignties. These issues are crucial for any advocacy and discussion around any policy document related to space sustainability and the impacts of satellite constellations.

The contributor Hilding Neilson (he/him) acknowledges that he is Mi'kmaq and a member of the Qalipu First Nation in Turtle Island (North America). His contribution is based on that perspective and based on numerous works of interdisciplinary research in astronomy, Indigenous studies, and research ethics. . 

I further acknowledge that any policy work relating to the field of astronomy benefits from historic and contemporary colonization. More specifically, we acknowledge that astronomers from around the globe and the International Astronomical Union have benefited from the taking of Indigenous lands for ground-based telescopes without free and informed consent or with manufactured consent.  This issue is particularly important today as Kanaka Maoli have spent decades protesting the Thirty Meter Telescope planned for construction on Mauna Kea in Hawai'i.  But, astronomers have benefited from the construction and planned construction of optical and radio facilities on Indigenous Lands the world over. There is some irony to the astronomy committee advocating for Indigenous voices in the discussion of building a framework for sharing outer space while preferring new observatories over respecting Indigenous rights (\cite{Prescod-Weinstein2020, Neilson2019, Kahanamoku2020, Salazar2014}; and more).

Even in the light of these issues of colonization that support Astronomy communities, the goals of this section are:
\begin{enumerate}
\item to highlight Indigenous rights with respect to outer space activities through the United Nations Declaration of the Rights of Indigenous Peoples (UNDRIP) \footnote{https://www.un.org/esa/socdev/unpfii/documents/DRIPS\_en.pdf},
\item to highlight the need for anti-colonial methods when parties advocate for new rules and regulations with respect to outer space activities,
\item to discuss how Indigenous methodologies can allow for novel regulations and perspectives on how humanity and private and public interests interact with outer space.
\end{enumerate}

Indigenous peoples have lived in regions around the world since time immemorial and continue to live in their territories or elsewhere today.  Many peoples have been removed from their territories because of various forms of genocides, but their rights and knowledges exist today and will continue to do so into the future.  As such, it is necessary for any policy analysis that wishes to be inclusive and equitable to be respectful of Indigenous knowledges. It is also necessary to note that there is no one Indigenous knowledge; every Nation or group has its own knowledge and its own methodologies.  Therefore, to be inclusive of Indigenous communities there must be consultations with as many different communities, and nations as possible.

\section{Indigenous rights to outer space via UNDRIP}

UNDRIP was adopted by the UN General Assembly in 2007 and has become a pathway for reconciliation in many nations.  In particular, for any discussion about outer space, we cite a number of articles that establish the rights of Indigenous peoples.

The noted articles include:
\begin{description}
\item[Article 1: ] Indigenous peoples have the right to the full enjoyment, as a collective or as individuals, of all human rights and fundamental freedoms as recognized in the Charter of the United Nations, the Universal Declaration of Human Rights and international human rights law. 
\item[Article 3: ] Indigenous peoples have the right to self-determination. By virtue of that right they freely determine their political status and freely pursue their economic, social and cultural development. 
\item[Article 5: ] Indigenous peoples have the right to maintain and strengthen their distinct political, legal, economic, social and cultural institutions, while retaining their right to participate fully, if they so choose, in the political, economic, social and cultural life of the State. 
\item[Article 7.1: ] Indigenous individuals have the rights to life, physical and mental integrity, liberty and security of person. 
\item[Article 8.1: ] Indigenous peoples and individuals have the right not to be subjected to forced assimilation or destruction of their culture. 
\item[Article 11.1: ] Indigenous peoples have the right to practice and revitalize their cultural traditions and customs. This includes the right to maintain, protect and develop the past, present and future manifestations of their cultures, such as archaeological and historical sites, artefacts, designs, ceremonies, technologies and visual and performing arts and literature. 
\item[Article 11.2: ] States shall provide redress through effective mechanisms, which may include restitution, developed in conjunction with indigenous peoples, with respect to their cultural, intellectual, religious and spiritual property taken without their free, prior and informed consent or in violation of their laws, traditions and customs. 
\item[Article 12.1: ] Indigenous peoples have the right to manifest, practice, develop and teach their spiritual and religious traditions, customs and ceremonies; the right to maintain, protect, and have access in privacy to their religious and cultural sites; the right to the use and control of their ceremonial objects; and the right to the repatriation of their human remains. 
\item[Article 13.1: ] Indigenous peoples have the right to revitalize, use, develop and transmit to future generations their histories, languages, oral traditions, philosophies, writing systems and literatures, and to designate and retain their own names for communities, places and persons. 
\item[Article 18: ] Indigenous peoples have the right to participate in decision-making in matters which would affect their rights, through representatives chosen by themselves in accordance with their own procedures, as well as to maintain and develop their own indigenous decision-making institutions. 
\item[Article 20.1: ] Indigenous peoples have the right to maintain and develop their political, economic and social systems or institutions, to be secure in the enjoyment of their own means of subsistence and development, and to engage freely in all their traditional and other economic activities. 
\item[Article 20.2: ] Indigenous peoples deprived of their means of subsistence and development are entitled to just and fair redress. 
\item[Article 25: ] Indigenous peoples have the right to maintain and strengthen their distinctive spiritual relationship with their traditionally owned or otherwise occupied and used lands, territories, waters and coastal seas and other resources and to uphold their responsibilities to future generations in this regard. 
\item[Article 26.1: ] Indigenous peoples have the right to the lands, territories and resources which they have traditionally owned, occupied or otherwise used or acquired. 
\item[Article 27: ] States shall establish and implement, in conjunction with indigenous peoples concerned, a fair, independent, impartial, open and transparent process, giving due recognition to indigenous peoples' laws, traditions, customs and land tenure systems, to recognize and adjudicate the rights of indigenous peoples pertaining to their lands, territories and resources, including those which were traditionally owned or otherwise occupied or used. Indigenous peoples shall have the right to participate in this process. 
\item[Article 29.1: ] Indigenous peoples have the right to the conservation and protection of the environment and the productive capacity of their lands or territories and resources. States shall establish and implement assistance programmes for indigenous peoples for such conservation and protection, without discrimination. 
\item[Article 29.2: ]States shall take effective measures to ensure that no storage or disposal of hazardous materials shall take place in the lands or territories of indigenous peoples without their free, prior and informed consent. 
\item[Article 31.1: ] Indigenous peoples have the right to maintain, control, protect and develop their cultural heritage, traditional knowledge and traditional cultural expressions, as well as the manifestations of their sciences, technologies and cultures, including human and genetic resources, seeds, medicines, knowledge of the properties of fauna and flora, oral traditions, literatures, designs, sports and traditional games and visual and performing arts. They also have the right to maintain, control, protect and develop their intellectual property over such cultural heritage, traditional knowledge, and traditional cultural expressions. 
\item[Article 32.1: ] Indigenous peoples have the right to determine and develop priorities and strategies for the development or use of their lands or territories and other resources. 
\item[Article 32.2: ] States shall consult and cooperate in good faith with the indigenous peoples concerned through their own representative institutions in order to obtain their free and informed consent prior to the approval of any project affecting their lands or territories and other resources, particularly in connection with the development, utilization or exploitation of mineral, water or other resources. 
\item[Article 32.3: ] States shall provide effective mechanisms for just and fair redress for any such activities, and appropriate measures shall be taken to mitigate adverse environmental, economic, social, cultural or spiritual impact. 
\item[Article 42: ] The United Nations, its bodies, including the Permanent Forum on Indigenous Issues, and specialized agencies, including at the country level, and States shall promote respect for and full application of the provisions of this Declaration and follow up the effectiveness of this Declaration.
\end{description}
 
All of these articles have some impact on Indigenous rights with respect to outer space even though the phrase ``outer space'' and this list, while detailed is not meant to be exhaustive.  We can consider these articles as addressing three different themes: 1) Cultural rights with respect to  outer space; 2) Economic fights with respect to outer space; and 3) Responsibilities with respect to outer space. It is in these three categories where Indigenous rights can be elucidated in more detail.

\section{Cultural rights and outer space}
Indigenous Peoples have lived in their territories since time immemorial. As such Indigenous Peoples have lived with the night sky in various ways also since time immemorial and those relationships manifest in numerous ways.  For instance, the Cree peoples on Turtle Island (North America) refers to themselves as Star People and as a people who come from the stars (\cite{Buck}).  What it means to come from the stars is concept that only Cree Elders and Knowledge Keepers can truly explain, but the concept does illustrate an intimate and relational connection with the night sky.  These relationships are apparent globally through Indigenous star stories and constructs such as Medicine Wheels, methods for navigating waters using the stars, calendars, and more. Regardless of how non-Indigenous persons might judges the values of these knowledges, i.e., wayfinding and navigation versus navigation with the Global Positioning System, the rights of Indigenous peoples to maintain, use, and live their cultural practices is part of rights from UNDRIP, in particular Article 31.1.  As such, any actions that remove access to the night sky for ritual, culture, etc., is a violation of UNDRIP.

Two ways that Indigenous peoples can lose access the night sky are: light pollution and satellite pollution.  \cite{Hamacher} have noted that light pollution is a form of cultural genocide; that is, if Indigenous stories are written in the stars, or in the case of some Nations, stories are written in the dark spaces between stars then removing visibility to those constellations or ``dark constellations'' is colonization and assimilation.  This is particularly true as Indigenous star knowledge is transferred as part of oral traditions; then, light pollution is a direct assault on Indigenous knowledges. This is also true for satellite pollution. Bright satellites (in this case bright does not necessarily have to mean visible by the unaided eye) that can be seen in various ways also impact how Indigenous peoples interact with the night sky.  It is unlikely that light pollution from satellite  onstellations will ever have a significant impact on Indigenous methods of wayfinding or time-keeping, but a bright satellite constellations will impact cultural relationships with the night sky and require Indigenous peoples to adapt their knowledges to a view of the Night Sky changed by humans without their consultation or consent. That, too, is colonization.

The ongoing impacts of light pollution and satellite pollution are a clear violation of Indigenous rights if they negatively impact the ability of peoples to transmit their knowledges and continue religious and spiritual practices (Articles 12.1 and 13.1). A lack of consent can be construed as a violation of Article 8 that demands that Indigenous peoples not be forced to assimilate. As such, any regulations and policies with respect to human interaction with outer space and the night sky require consultation and consent of Indigenous peoples from the world over.

\section{Economic rights and outer space} 
Many of the Articles of UNDRIP highlight the rights of Indigenous peoples to manage their economic well-being and the right to access their traditional territories for development. This includes: Articles 3; 8, and 32.  The rights of Indigenous peoples to control economic development in outer space on Indigenous lands can be considered through two lenses: 1) from the usage of lands for launch capabilities and from the usage of outer space itself.
There continues to be significant interest in creating more launch facilities to expedite human activity in space.  Many of these launch facilities are being built on traditional Indigenous lands without free and prior consent and without economic benefit for the peoples living on those lands. As launch facilities can contribute significant pollution via noise, and blowing dust around, and in waste if launches fail, then there is an ethical, if not, legal, obligation, for space companies to negotiate and consider the rights of Indigenous peoples.

As for outer space itself, it may seem counterintuitive to consider that Indigenous peoples would have any more economic rights than any other group on Earth.  However, the UNDRIP guarantees the rights of Indigenous peoples to use their traditional lands.  Furthermore, traditional lands do not have to be defined solely by the colonial definitions of borders as defined by Articles 26 and 27. Those articles define lands in terms of ``traditionally owned, occupied or otherwise used or acquired''.  Historically, nation states may have defined traditional lands in very strict terms of the written records of colonizing nations and by archaeological evidence; i.e., did Indigenous peoples leave physical evidence on specific spots of land.  These methods were specifically designed to dispossess Indigenous peoples of traditional lands and do not reflect Indigenous claims and oral knowledges. As such, the rights and responsibilities of Indigenous peoples with respect to land should respect any Indigenous groups' definition of Land which would include access to all components that contribute to the ecosystem and culture such as water, air, the night sky, and yes, outer space. 

These rights are enshrined in many oral traditions. In North America alone, many First Nations peoples refer to themselves as coming from the stars, for which it is not within the rights of anyone from outside those cultures to define what ``coming from the stars'' mean.  Many stories include people coming and going from outer space, making Indigenous peoples ``users'' of and visitors to outer space.  Denying rights to Indigenous peoples to be part of the regulation of how humans operate in outer space because there is no evidence that they use outer space or used technology to go there is to perpetuate colonization.

As such, it is reasonable to argue that celestial objects, such as the Moon, and outer space, itself, are part of the traditional lands of Indigenous peoples and in the framework of UNDRIP means that Indigenous peoples have the right to participate economically as well as to have a say in how the outer space environment is regulated.

\section{The obligation of the United Nations and space actors}
Because Indigenous peoples have rights to participate in the space economy and to have cultural rights to space protected  as they choose, it is then incumbent on the United Nations and nation states to uphold UNDRIP and work with Indigenous peoples in developing the path for humanities present and future in outer space. There are a number of ways to approach this collaboration and consultation which include:
\begin{itemize}
\item  Organizations choosing to engage in dialogues about outer space regulation must consult with and equitably include Indigenous peoples in that work.  It is not sufficient to include Indigenous peoples solely through simplistic lenses of `Cultural Engagement' or `Stakeholders'.
\item  Nation states must consult and work with Indigenous peoples, in the framework of free and prior informed consent (FPIC) to develop an anti-colonial framework to outer space activities that moves beyond the simple colonial framework that has traditionally existed.
\item  Nation states must consult and work with Indigenous peoples to support their ability to benefit from the space economy in as equitable of ways as other actors.  
\item  The United Nations, in developing new regulations for the use of outer space, must consult and work with Indigenous peoples under the framework of free and prior informed consent in ways that acknowledge Indigenous peoples as nations unto themselves and respects their sovereignties.
\end{itemize}
These recommendations do not have a clear basis in international law, but are anti-colonial methods to support the rights of Indigenous peoples. They are a starting point for developing a new anti-colonial framework for our interactions with outer space and celestial objects in ways that acknowledge and respect Indigenous sovereignties in their respective lands and respects both their cultural and economic rights.

\section{Indigenous rights and sustainability of outer space}
One of the ways we might consider supporting Indigenous cultural and economic rights in outer space would be moving to a paradigm that goes beyond traditional settler perspectives of sustainability that are arguably framed around the issue of exploitation of resources and how much impact on the environment is permissible before the damage to the environment is irreversible or human usage of that environment is unsafe.  Historically, Indigenous peoples have suffered under this kind of discussion.  The nuclear tests of the Manhattan Project impact Navajo peoples, while nuclear bomb testing in the Pacific continues to have negative health impacts on peoples there.  Uranium mining has harmed both Navajo and Inuit communities.  Despite all of that, many consider nuclear power to be a sustainable and environmentally friendly method of electricity generation.

There are many more examples of how the history and context of environmentalism and sustainability has not included Indigenous peoples nor respected their land rights or sovereignties. The current global context of the Outer Space Treaty and the Artemis Accords is based on how humans, or more specifically nation states, operate with respect to one another in outer space. These agreements have no context for how humans or nation states or private companies operate with respect to outer space. Effectively, these agreements, more so the Artemis Accords, focus on ways to peacefully exploit outer space and protect the ``rights'' of those doing so.

One way to consider our interaction in space is via a treaty, but not a treaty about human activities in outer space, but a treaty about our responsibilities in outer space. Many Indigenous peoples traditionally lived in treaty and relationship with animals, plants, water, air, etc.  These treaties can and have included outer space. In the Mi'kmaw cosmology of six worlds, the Land above Clouds, or oouter pace, is part of a treaty.  \cite{SkyCountry} notes that Sky Country (Space) is to be cared for, not exploited. These treaties do not outline rights to outer space, but ethical obligations and responsibilities.  

It is reasonable that any anti-colonial perspective on outer space not only include the rights of Indigenous peoples, but also the rights of celestial objects and outer space itself and thus humanity's responsibilities for engaging in activities in outer space.  The ways this would work would not be obvious, especially in a global discussion that involves colonial nation states. However, many scholars have noted that responsibilities can include giving back in trade when using resources  This option requires humanity to operate in a system that is not focused solely on exploitation but supporting an ecosystem at the same time.  In outer space, the concept of an ecosystem would be difficult to understand for many because in the Western tradition, outer space and celestial objects are treated as dead/non-living, and in the Western system where humans are the apex of a hierarchy, humans would have a right to exploit however they see fit. For many Indigenous peoples these treaties with land, animals, and space mean that they are animate or living. As such any resource usage must be in trade and must benefit space or the Moon. 

One might question how we can determine what would be of benefit for space or celestial objects and who would determine the answer to that question.  It is not obvious how to answer this question because any international committee to address such a question can have dramatically different opinions and powers, but options might include if we take a resource on the Moon, then we have to return it. If we use water on the Moon for sustaining settlements or for propulsion then we might have an obligation to return an equal amount of water to the Moon. This concept is a way to maintain sustainability in a way that is not common on Earth but will also require changes to how we relate to life on Earth.  If light pollution is treated this way, then we have an obligation to the animals that are impacted by light pollution and have to give back to them in ways that support them. 

This concept is simply a different perspective on much of the current international discussion that is, by and large, ignoring the rights and voices of Indigenous peoples in favour of various colonial and capitalist interests. But, if we choose to follow UNDRIP and work with Indigenous nations the world over then there may be more options than the bimodal and colonial discussion of exploitation and sustainability.

\end{document}